\documentclass[prl,aps,twocolumn,superscriptaddress]{revtex4}
\usepackage{psfrag,slashed,cancel,array,graphicx,todonotes,hyperref}
\usepackage[utf8]{inputenc}
\usepackage{mathtools}
\usepackage{mathrsfs}
\usepackage{multirow}
\usepackage{braket}
\usepackage{slashed}
\usepackage{booktabs} 
\usepackage[normalem]{ulem}
\usepackage{slashed}
\hypersetup{pdftitle={},pdfcreator={},linkcolor=[rgb]{0.15,0.35,0.75},colorlinks=true,citecolor=[rgb]{0.675,0,0.2},urlcolor=[rgb]{0.15,0.35,0.65}}
\allowdisplaybreaks

\def\beq{\begin{equation}}
\def\eeq{\end{equation}}
\def\bsp#1\esp{\begin{split}#1\end{split}}
\def\d{{\rm d}}

\newcommand{\nn}{\nonumber}

\AtBeginDocument{
\heavyrulewidth=.08em
\lightrulewidth=.05em
\cmidrulewidth=.03em
\belowrulesep=.65ex
\belowbottomsep=0pt
\aboverulesep=.4ex
\abovetopsep=0pt
\cmidrulesep=\doublerulesep
\cmidrulekern=.5em
\defaultaddspace=.5em
}

\def\be{\begin{equation}}
\def\ee{\end{equation}}

\begin{document}

\preprint{XXX}

\author{Liang Dong}
\email{liang.dong@sjtu.edu.cn}
\affiliation{State Key Laboratory of Dark Matter Physics, Shanghai Key Laboratory for Particle Physics and Cosmology, Key Laboratory for Particle Astrophysics and Cosmology (MOE),
School of Physics and Astronomy, Shanghai Jiao Tong University, Shanghai 200240, China}
\author{Shen Fang}
\email{sfang23@m.fudan.edu.cn}
\affiliation{Department of Physics, Center for Field Theory and Particle Physics, and Key Laboratory of Nuclear Physics and Ion-beam Application (MOE), Fudan University, Shanghai, 200433, China}
\author{Jun Gao}
\email{jung49@sjtu.edu.cn}
\affiliation{State Key Laboratory of Dark Matter Physics, Shanghai Key Laboratory for Particle Physics and Cosmology, Key Laboratory for Particle Astrophysics and Cosmology (MOE),
School of Physics and Astronomy, Shanghai Jiao Tong University, Shanghai 200240, China}
\author{Hai Tao Li}
\email{haitao.li@sdu.edu.cn}
\affiliation{School of Physics, Shandong University, Jinan, Shandong 250100, China}
\author{\\  Ding Yu Shao}
\email{dyshao@fudan.edu.cn}
\affiliation{Department of Physics, Center for Field Theory and Particle Physics, and Key Laboratory of Nuclear Physics and Ion-beam Application (MOE), Fudan University, Shanghai, 200433, China}
\affiliation{Shanghai Research Center for Theoretical Nuclear Physics, NSFC and Fudan University, Shanghai 200438, China}
\affiliation{Center for High Energy Physics, Peking University, Beijing 100871, China}
\affiliation{Southern Center for Nuclear-Science Theory (SCNT), Institute of Modern Physics, Chinese Academy of Sciences, Huizhou 516000, Guangdong Province, China}
\author{Hua Xing Zhu}
\email{zhuhx@pku.edu.cn}
\affiliation{School of Physics, Peking University, Beijing 100871, China}
\affiliation{Center for High Energy Physics, Peking University, Beijing 100871, China}
\affiliation{Southern Center for Nuclear-Science Theory (SCNT), Institute of Modern Physics, Chinese Academy of Sciences, Huizhou 516000, Guangdong Province, China}
\author{Yu Jiao Zhu}
\email{yzhu@mpp.mpg.de}
\affiliation{Max-Planck-Institut f\"ur Physik, Werner-Heisenberg-Institut, Boltzmannstrasse 8, 85748 Garching, Germany}

\title{Two-Dimensional Transverse-Momentum Subtraction and Semi-Inclusive Deep-Inelastic Scattering at N$^3$LO in QCD}

\begin{abstract}
Identified hadron production is essential for the study of nucleon structure and QCD hadronization at high energies. 
We present the \textit{first} calculation of unpolarized semi-inclusive deep-inelastic scattering (SIDIS) at next-to-next-to-next-to-leading order (N$^3$LO) in perturbative QCD. 
Our calculation is based on a novel method of \textit{two-dimensional} transverse-momentum subtraction motivated by QCD factorization of soft and collinear singularities.  
The N$^3$LO corrections are moderate in general but can be significant in threshold regions, and exhibit excellent perturbative convergence and reduced scale variations.
The fully differential framework allows for arbitrary selection cuts and directly enables precision nucleon tomography at the upcoming Electron-Ion Collider, establishing the theory foundation needed to match the anticipated experimental accuracy. 
Generalization of the method to calculations of polarized SIDIS is also feasible.   
\end{abstract}

\maketitle

\paragraph*{Introduction.---}
Precision predictions in perturbative Quantum Chromodynamics~(QCD) form the backbone of the collider physics program, essential for exploiting the physics potential of the Large Hadron Collider and future collider facilities. 
Driven by innovative developments in theoretical methods and computational techniques, the frontier of perturbative QCD calculations has advanced to next-to-next-to-next-to-leading order (N$^3$LO) for several benchmark processes, including Higgs boson production~\cite{Anastasiou:2015vya, Anastasiou:2016cez, Dreyer:2016oyx, Mistlberger:2018etf, Duhr:2019kwi, Duhr:2020kzd, Baglio:2022wzu}, Drell-Yan lepton pair production~\cite{Duhr:2020seh, Duhr:2020sdp, Duhr:2021vwj, Chen:2022cgv}, Higgs boson pair production~\cite{Chen:2019fhs, Chen:2019lzz}, and top-quark decay~\cite{Chen:2023osm, Chen:2023dsi}. 
However, these milestones have largely been restricted to processes without final-state color flow at the Born level, or inclusive over final states.
Recently, the precision calculation for identified hadron production has garnered significant attention~\cite{Abele:2021nyo, Goyal:2023zdi, Bonino:2024qbh, Bonino:2024wgg, Goyal:2024tmo, Goyal:2024emo, Bonino:2025tnf, Haug:2025ava, Bonino:2025qta, Zhou:2025lqv, Bonino:2025bqa, Goyal:2025qyu, Goyal:2025bzf}. These processes provide a unique window into the confinement mechanism, allowing for the rigorous extraction of non-perturbative fragmentation functions~\cite{Bertone:2017tyb, Soleymaninia:2018uiv, Moffat:2021dji, AbdulKhalek:2022laj, Borsa:2022vvp, Gao:2024nkz, Gao:2024dbv, Gao:2025hlm, Gao:2025kof, Gao:2025bko}, the probing of the proton's spin structure~\cite{Bertone:2024taw, Borsa:2024mss, Cruz-Martinez:2025ahf, Cocuzza:2025qvf}, and the search for flavor-dependent new physics.
From a broader perspective, identifying hadrons within the final state is a requisite for understanding the parton-to-hadron transition. 
In particular, in the small-angle factorization limit, the production of identified hadrons serves as a critical theoretical ingredient for describing the internal structure of jets~\cite{Procura:2009vm, Jain:2011xz, Chang:2013rca, Ritzmann:2014mka, Kaufmann:2015hma, Dai:2016hzf, Chien:2016led, Kang:2016ehg, Kang:2017glf, Anderle:2017cgl, Kang:2019ahe, Kang:2021ffh, Caletti:2022glq} and calculating energy correlator observables~\cite{Dixon:2019uzg, Chen:2020vvp}. 
Despite this importance, theoretical progress has been hindered by the complexity of handling final-state collinear singularities associated with the observed hadron. While calculations have matured at next-to-next-to-leading order (NNLO)~\cite{Fu:2024fgj, Czakon:2025yti, Generet:2025bqx, Gao:2026tnd, Bonino:2026dvr, Dong:2026eas}, the N$^3$LO frontier remains unexplored, with the sole exception of single-inclusive annihilation, $e^+e^- \to h +X$~\cite{He:2025hin}.
The experimental landscape for identified hadron production is already rich, with abundant data available from $e^+e^-$, $ep$, and $pp$ collisions.
Looking forward, future programs such as the Electron-Ion Collider (EIC)~\cite{AbdulKhalek:2021gbh, AbdulKhalek:2022hcn} and future $e^+e^-$ factories~\cite{CEPCStudyGroup:2018ghi, FCC:2018evy} promise measurements of unprecedented precision. 
To fully exploit the potential of these datasets and match the experimental accuracy, it is imperative to elevate the theoretical precision of these probes to the N$^3$LO level.

\begin{figure*}[t!]
    \centering
    \includegraphics[width=0.45\textwidth]{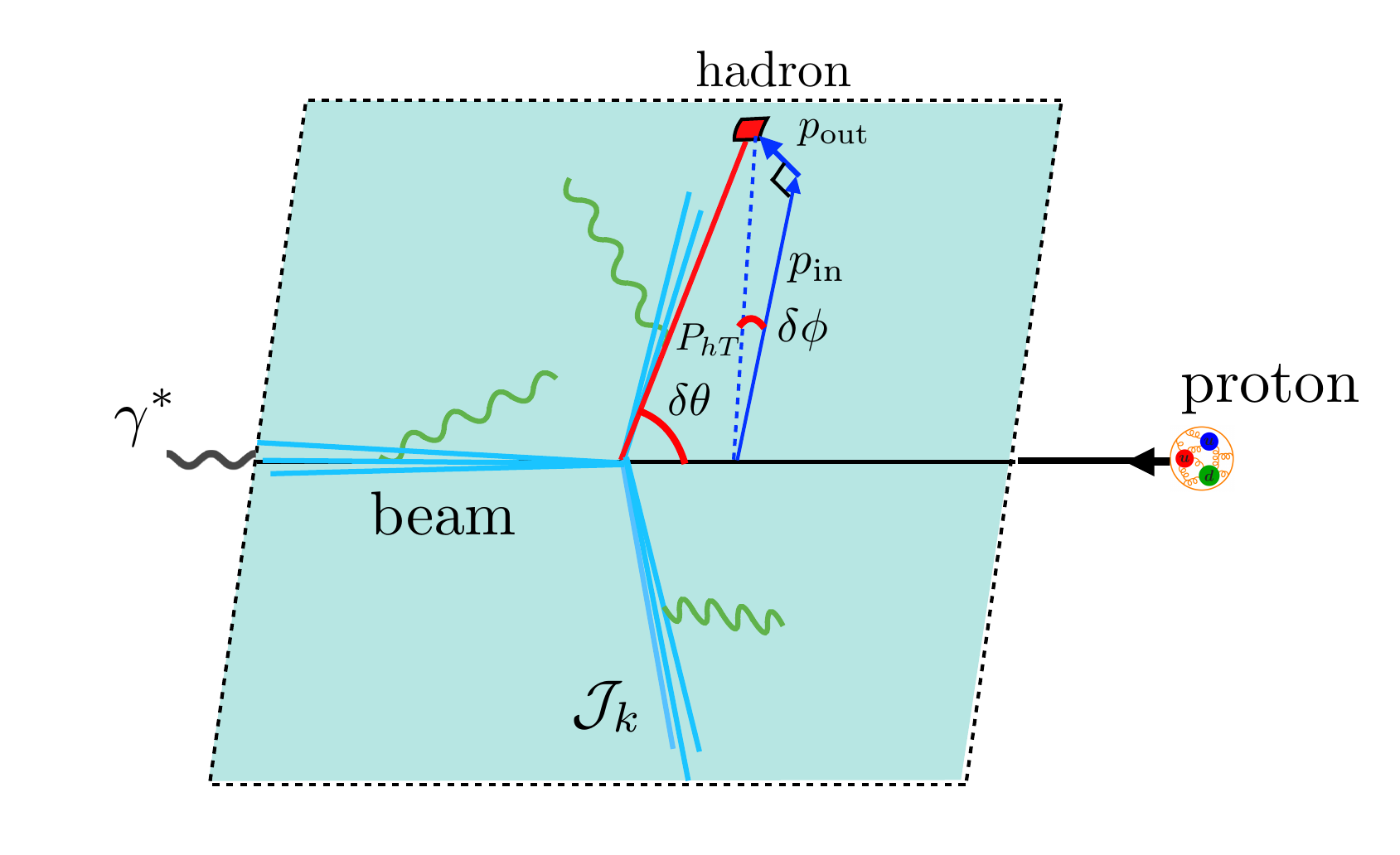}
    \includegraphics[width=0.38\textwidth]{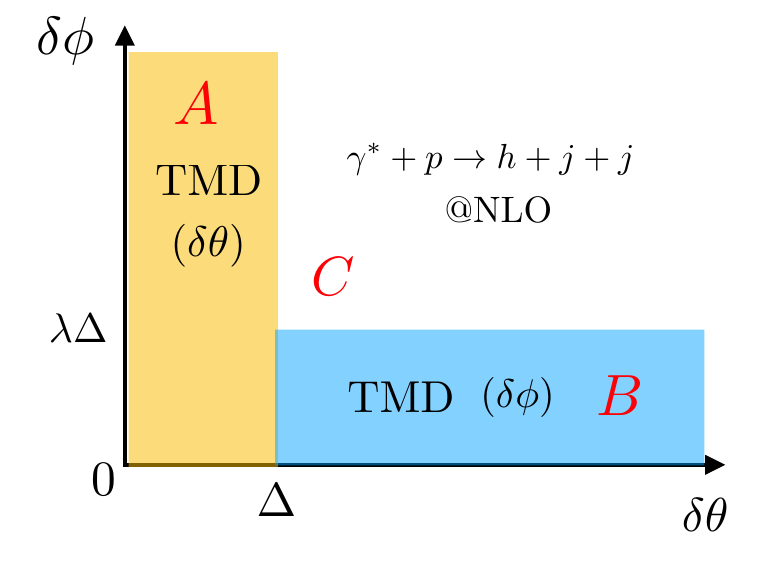}
    \vspace{-0.5cm}
    \caption{
    Left: schematic definition of the kinematic variables used for the two-dimensional $q_T$-subtraction method in the Breit frame. 
    The incoming virtual photon $\gamma^*$ scatters off the proton, producing an identified hadron and a recoiling leading jet ${\cal J}_k$. 
    The blue lines along the proton direction denote beam collinear radiation. 
    The green curved lines indicate the soft radiation.  
    The slicing variables are chosen as the beam polar angle $\delta \theta$ and the azimuthal decorrelation $\delta\phi$.
    Right: partition of the phase space into different regions according to $\delta \theta > \Delta$ and $\delta \phi > \lambda \Delta$, with the two slicing parameters $\Delta,\,\lambda \ll 1$. 
    }
\label{fig:kinematics}
\end{figure*}

In this Letter, we propose a novel approach of \textit{two-dimensional} transverse-momentum subtraction
to describe the production of identified hadrons in $ep$ and $e^+e^-$ collisions, generalizing the $q_T$-subtraction method for inclusive observables~\cite{Catani:2007vq,Catani:2009sm,Catani:2010en,Catani:2011qz}.
Our framework relies on a rigorous understanding of the infrared structure of gauge theories~\cite{Fu:2024fgj, Gao:2026tnd, Dong:2026eas}, utilizing the formalism of soft-collinear effective theory (SCET)~\cite{Bauer:2000yr, Bauer:2001ct, Bauer:2001yt}.
As a concrete application, we present the \textit{first} N$^3$LO calculation for the semi-inclusive deep-inelastic scattering (SIDIS) process, $e^- + p \to e^- + h + X$.
Beyond this specific process, our results also provide the necessary perturbative ingredients to compute identified hadron production within jets~\cite{Bonino:2026dvr} at N$^3$LO.

\paragraph*{The method.---}
In the Breit frame, the scattering kinematics of the SIDIS process, $\gamma^*+p \to h+X$, are illustrated in the left panel of Fig~\ref{fig:kinematics}.
At leading order (LO), the hadron is produced with zero transverse momentum $P_{hT}$, or equivalently with beam polar angle $\delta\theta =0$, where $P_{hT}\equiv P_h \sin\delta\theta$. 
Beyond LO, the identified hadron acquires a transverse momentum $\vec{P}_{hT}=(p_{\rm in},p_{\rm out})_J$, consisting of components within and perpendicular to the scattering plane defined by the incoming beams and the recoiling leading jet with the winner-take-all (WTA) recombination scheme~\cite{Salam:WTAUnpublished, Bertolini:2013iqa}. 
They can be parametrized by the azimuthal decorrelation $\delta \phi$, with $p_{\rm in}\equiv P_{hT}\cos\delta\phi$ and $p_{\rm out}\equiv P_{hT}\sin\delta\phi$. 
Consequently, at next-to-leading order (NLO) the hadron retains a vanishing out-of-plane momentum $p_{\rm out}$, or equivalently, $\delta \phi=0$.
Thus, by requiring both non-vanishing $P_{hT}$ and $p_{\rm out}$, the soft and collinear pattern of QCD radiation can be largely simplified.  
Especially, the N$^3$LO corrections in this region are entirely captured by the NLO calculation of the associated production of the hadron and a dijet. 
Motivated by these kinematic boundaries, we propose a two-dimensional transverse-momentum subtraction scheme governed by two slicing variables, $P_{hT}$ and $p_{\rm out}$ (or equivalently, $\delta\theta$ and $\delta\phi$).
Following the $q_T$-subtraction approach for identified hadron production~\cite{Gao:2026tnd, Fu:2024fgj, Dong:2026eas}, we partition the cross section as
\begin{align}\label{eq:full_xsection}
     \frac{\d\sigma}{\d\mathcal{O}}  & = \int_{0}^{\Delta} \d \delta\theta  \frac{\d\sigma^A}{ \d\delta\theta {\d\mathcal{O}} } + \int_{\Delta}^{\delta\theta ^{\rm max}} \d \delta\theta  \\
     &\times\Big( \int_{0}^{\lambda \Delta} \d \delta\phi \frac{\d\sigma^{B}}{ \d \delta\theta \d\delta\phi{\d\mathcal{O}} } 
      + \int_{\lambda \Delta}^{\delta\phi^{\rm max}} \d \delta\phi \frac{\d\sigma^{C}}{ \d \delta\theta \d\delta\phi{\d\mathcal{O}} }\Big)\nn,
\end{align}
where the three terms correspond to contributions from regions $A$, $B$, and $C$, respectively, as shown in the right panel of Fig~\ref{fig:kinematics}.
The contributions from region $A$ lead to a tree-level kinematics, and we have suppressed the dependence on $\delta\phi$ which has been integrated over in later calculations.
The boundaries are characterized by two small resolution or slicing parameters, $\Delta$ and $\lambda \ll 1$.
For the resolved region $C$, the NLO calculations are carried out based on the {\tt FMNLO} framework~\cite{Liu:2023fsq} for identified hadron production.
The necessary one-loop matrix elements for $ei \to ejkl$ and tree-level matrix elements for $ei \to ejklm$ are taken from Refs.~\cite{Bern:1997sc, Campbell:2002tg, Campbell:2003hd, Campbell:2010ff}, where $ijklm$ represent all possible flavor combinations of QCD partons. 
Concerning the two unresolved regions, for region $A$ in the limit $P_{hT}\ll Q$, the cross section is governed by the transverse-momentum-dependent (TMD) factorization formula~\cite{Gao:2026tnd}
\begin{align} \label{eq:sidis}
      &\frac{\d\sigma^A}{\d x \, \d y \, \d z \, \d^2 \vec P_{hT} }  \propto \int \frac{\d^2 \vec{b}_{\perp}}{4\pi^2} e^{-i \vec P_{hT} \cdot \vec{b}_\perp/z}  \sum_{i}   H_{ei\to ei}(Q)
      \nn\\
  &\hspace{0.5cm}\times  {\cal B}_{i/p}(x,\vec{b}_\perp)\,
      {\cal D}_{h/i}(z,\vec{b}_\perp) \, S_{qq}(\vec{b}_\perp )\left[1 + \mathcal{O}(\Delta )\right] 
\end{align}
at leading power, and can be calculated using its fixed-order expansions to desired orders of $\alpha_s$ with power corrections vanishing in the limit of $\Delta\to 0$. 
Here $Q$, $x$, $y$, and $z$ are the momentum transfer, Bjorken variable, inelasticity and identified hadron momentum fraction, respectively.
%
%
The hard function $H_{ei\to ei}$ is obtained from the quark matching coefficients extracted from the quark form factors~\cite{Becher:2006mr, Moch:2005tm, Moch:2005id, Baikov:2009bg, Gehrmann:2010ue, Gehrmann:2010tu, Lee:2022nhh}.
The $ {\cal B}_{i/p}$ and  ${\cal D}_{h/i}$ denote the standard TMD beam function for parton $i$ in proton $p$ and TMD fragmentation function for parton $i$ fragmenting into hadron $h$, respectively, and both of them are known through three loops~\cite{Luo:2019szz,Luo:2020epw,Ebert:2020yqt}. 
The three-loop analytic result for the soft function $S_{qq}$ can be obtained from \cite{Li:2016ctv}, thanks to the crossing property at this order~\cite{Zhu:2020ftr}.
\begin{figure*}[t!]
    \centering
    \includegraphics[width=0.92\textwidth]{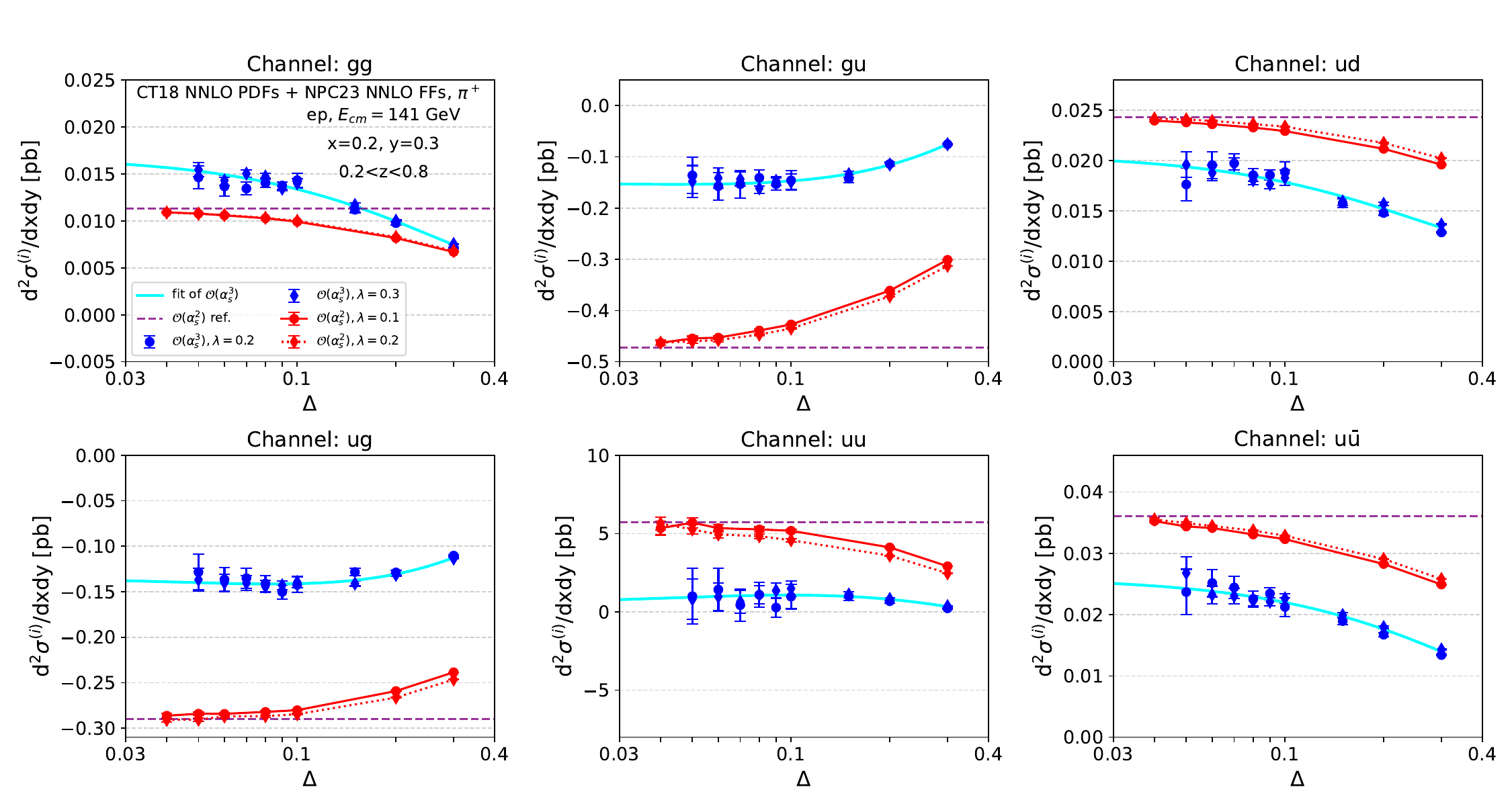}
    \vspace{-0.5cm}
    \caption{%
    Numerical stability of corrections to the total cross section with respect to the slicing parameter $\Delta$.
    The panels display the $\mathcal{O}(\alpha_s^2)$ (red) and $\mathcal{O}(\alpha_s^3)$ (blue) contributions for six representative partonic channels. 
    Error bars indicate Monte Carlo statistical uncertainties.
    The solid cyan lines represent a fit to the asymptotic limit $\Delta \to 0$. 
    The dashed purple lines denote the reference values of $\mathcal{O}(\alpha_s^2)$.}
    \label{fig:cutoff_dep}
\end{figure*}
Similarly, for region $B$ in the asymptotic limit where the out-of-plane momentum is small ($p_{\rm out} \ll P_{hT}$), the differential cross section is governed by the TMD factorization formula~\cite{Chien:2020hzh, Chien:2022wiq, Fu:2024fgj, Dong:2026eas, Fu:2026nkd}
\begin{align} \label{eq:fact_phi_jet}
      &\frac{\d\sigma^B}{\d x \, \d y \, \d z \, \d^2 \vec P_{hT}  \, \d p_{\rm out}} \propto
      \int \frac{\d b}{2\pi}e^{i p_{\rm out} b/\zeta} \\
  &\hspace{1cm}\times   \sum_{ijk} \int \d \xi\,   H_{ei\to ejk}(Q,\xi,\zeta)\nn\\
  &\hspace{1cm}\times  {\cal B}_{i/p}(\xi,b)\,
      {\cal D}_{h/j}(\zeta,b)\,{\cal J}_{k}(b)  \, S_{ij k}(b)\left[1 + \mathcal{O}(\lambda)\right] \nn
\end{align} 
at leading power, and can be calculated up to power corrections that vanish in the
limit $\lambda\to0$. 
Here, $\mathcal{J}_k$ denotes the TMD jet function for the recoiling jet initiated by parton $k$, defined using the $k_T$ algorithm and the WTA recombination scheme~\cite{Gutierrez-Reyes:2019vbx, Fang:2024auf, Bell:2021dpb, Brune:2022cgr, Buonocore:2025ysd}.
The function $S_{ijk}(b)$ is the TMD soft function associated with the colored directions $i$, $j$, and $k$~\cite{Echevarria:2015byo, Lubbert:2016rku,Gao:2019ojf, Chien:2020hzh, Chien:2022wiq, Gao:2023ivm, Fu:2024fgj}.

To achieve N$^3$LO accuracy for SIDIS, the $\mathcal O(\alpha_s^3)$ expansion of Eq.~(\ref{eq:fact_phi_jet}) is required, which was validated in ~\cite{Dong:2026eas}. 
For a detailed discussion of the kinematic variables ($\xi$ and $\zeta$), and the hard, soft, and jet functions, we refer the reader to Ref.~\cite{Dong:2026eas}.
After combining the contributions from regions $A$, $B$, and $C$, the leading-power dependence on $\Delta$ and $\lambda$ cancels, while residual power corrections remain that may be linear or quadratic depending on the fiducial cuts~\cite{Ebert:2018gsn, Camarda:2021jsw, Buonocore:2021tke}.
A dedicated study of these effects is beyond the scope of this work.
Furthermore, our two-dimensional subtraction method is equally applicable to hadron production in single-inclusive annihilation (SIA) of $e^+e^-$ collisions.
In that case, one simply replaces the proton with the sub-leading jet, which defines the scattering plane together with the leading recoiling jet.
In the factorization formulas, the beam function is replaced by another jet function.
A fully differential calculation of SIA at N$^3$LO will be complementary to the analytical results~\cite{He:2025hin}.

\paragraph*{Numerical results.--}

In this section we present phenomenological studies of SIDIS at N$^3$LO in QCD.
We consider electron-proton collisions at a center of mass energy of the EIC ($\sqrt s=141$ GeV) and use the CT18 NNLO parton distribution functions (PDFs)~\cite{Hou:2019efy} of five active quark flavors with the strong coupling constant $\alpha_s(M_Z)=0.118$.  
We calculate differential cross sections for the charged pion ($\pi^+$) production with NPC23 NNLO fragmentation functions (FFs)~\cite{Gao:2025hlm} through all orders, and use a nominal choice of $\mu=Q$ for both the renormalization and factorization scales. 
We have neglected contributions from the $Z$ boson exchange for simplicity, which are small for the energies considered.
To validate our calculation framework, we explicitly verify that the cross section remains independent of the arbitrary cutoff $\Delta$ in the limit $\Delta\to 0$. 
Fig.~\ref{fig:cutoff_dep} displays the numerical stability of the $\mathcal{O}(\alpha_s^2)$ (NNLO) and $\mathcal{O}(\alpha_s^3)$ (N$^3$LO) corrections to the differential cross section of charged pion production at the EIC, for two representative values of $\lambda$, 0.3 and 0.2 (0.2 and 0.1) at N$^3$LO (NNLO).
The kinematics are fixed at $x=0.2$, and $y=0.3$, with integration over $0.2<z<0.8$. 
The results are shown for six representative partonic channels, including $uu$, $ug$, $gu$, $gg$, $u\bar u$, and $u d$, labeled according to the flavors of the incoming and fragmenting partons.
Among them the $uu$ channel is dominant due to both favored PDFs and FFs. 
The NNLO results (red points) exhibit excellent flatness over $0.03 < \Delta< 0.1$, confirming a precise cancellation of infrared divergences, and converge to the reference values calculated in Refs.~\cite{Goyal:2024tmo, Bonino:2024qbh}. 
At N$^3$LO (blue points), the results similarly converge to a constant plateau, despite the larger Monte Carlo integration uncertainties characteristic of the resolved cross section.
Results from two different choices of $\lambda$ are consistent, given small values of $\lambda$.
The solid cyan curves in Fig.~\ref{fig:cutoff_dep} represent fits to the genuine $\mathcal{O}({\alpha_s^3})$ corrections including power corrections~\cite{Grazzini:2017mhc, Ebert:2019zkb} while the red lines are connections of the $\mathcal{O}({\alpha_s^2})$ data points. 
The explicit functional forms are given in the Supplemental Material~\cite{SupplementalMaterial}.
From the fit we decide that a choice of $\Delta=0.1$ in the N$^3$LO calculations is sufficient for keeping power corrections under control in a phenomenological study.
Using even smaller values of $\Delta$ will require extensive computational resources in order to maintain comparable numerical precision.  
The stability observed across both quark- and gluon-initiated channels establishes the robustness of our method.
Overall we find good convergence of the perturbative expansion with N$^3$LO corrections much smaller than NNLO corrections, especially for the three favored channels, $uu$, $ug$ and $gu$.
For $ud$, $u\bar u$ and $gg$ channels, which start entering at NNLO, the N$^3$LO corrections can be comparable to or even larger than the NNLO corrections. 
In the calculations, we have used approximate values of the two-loop constant terms for the WTA unpolarized and linearly polarized gluon jet functions $j_{g, \text{WTA}}^{(2)}$, which enter the N$^3$LO corrections of the $uu$ channel~\cite{Dong:2026eas}. 
They are estimated by assuming a scaling behavior between the energy-energy correlation (EEC) and WTA observables. 
Specifically, we apply the relative factor between the constant terms of the WTA quark $j_{q, \text{WTA}}^{(2)}$~\cite{Fang:2024auf} (see also Refs. \cite{Gutierrez-Reyes:2019vbx, Bell:2021dpb, Brune:2022cgr, Buonocore:2025ysd}) and the EEC $j_{q, \text{EEC}}^{(2)}$~\cite{Luo:2019hmp} jet functions to the known gluon EEC results $ j_{g, \text{EEC}}^{(2)}$~\cite{Luo:2019bmw} as follows:
\begin{align}
    j_{g_U, \text{WTA}}^{(2)} &\approx j_{g_U, \text{EEC}}^{(2)} \times \frac{j_{q, \text{WTA}}^{(2)}}{j_{q, \text{EEC}}^{(2)}}=-12.9\,, \nn\\
    j_{g_L, \text{WTA}}^{(2)} &\approx j_{g_L, \text{EEC}}^{(2)} \times \frac{j_{q, \text{WTA}}^{(2)}}{j_{q, \text{EEC}}^{(2)}}=0.97\,, 
\end{align}
where $U$ and $L$ denote the unpolarized and linearly polarized gluons, respectively.
We have checked that, when increasing the above unpolarized (polarized) value to two times, the N$^3$LO corrections to the $uu$ channel change by less than 2\% (0.2\%) of the NNLO corrections.
Meanwhile, the value of $ j_{g_U, \text{WTA}}^{(2)}$ can be extracted utilizing decays of the Higgs boson into two gluons.
We found a result of $-9.4\pm 3.2$ by comparing Monte Carlo (MC) calculations~\cite{Hu:2021rkt} to the known partial width at NNLO~\cite{Baikov:2006ch}, which is consistent with our estimation. 
\begin{figure}[t!]
    \includegraphics[width=0.9\linewidth]{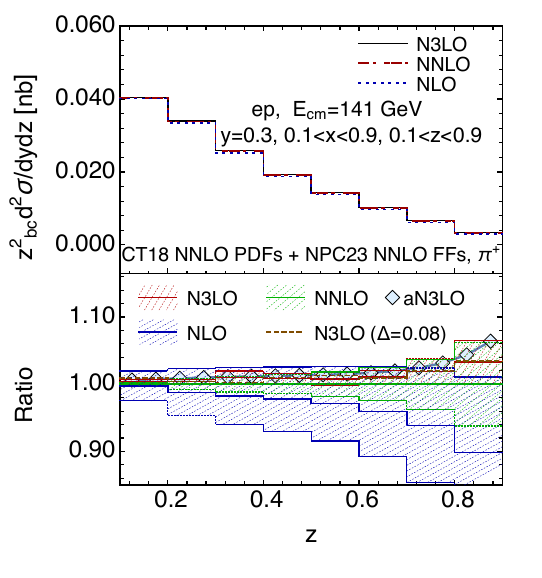}
    \vspace{-0.5cm}
    \caption{%
    {Differential cross section (multiplied by square of bin center) as a function of the hadron momentum fraction $z$ for charged pion production at the EIC at various orders in QCD, including NLO (blue dotted), NNLO (red dot-dashed) and N$^3$LO (black solid).
    In the lower panel, all predictions including scale variations are normalized to the central NNLO results.
    The brown dashed curve shows the N$^3$LO prediction for the alternative slicing cutoff $\Delta=0.08$; the default is $\Delta=0.1$.
    The diamond markers show the approximate N$^3$LO predictions, and the lines connecting them are included only as guides to the eye.}
    }
    \label{fig:z_dist}
\end{figure}
\begin{figure}[t!]
    \includegraphics[width=0.9\linewidth]{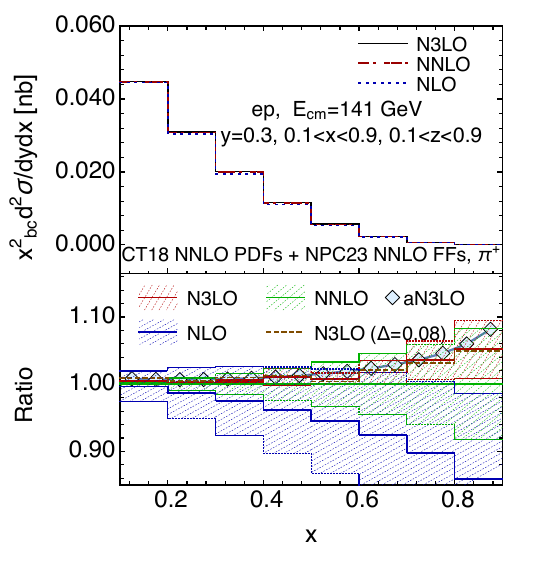}
    \vspace{-0.5cm}
    \caption{%
   {
   Charged pion differential cross section at the EIC as a function of the Bjorken variable $x$.
   The perturbative orders, normalization, scale variations, slicing-cutoff choices, and approximate N$^3$LO markers follow the conventions of Fig.~\ref{fig:z_dist}.
   }
    }
    \label{fig:x_dist}
\end{figure}
Fig.~\ref{fig:z_dist} presents our N$^3$LO predictions for the charged pion differential cross section as a function of the momentum fraction $z$, tailored to the kinematics of the future EIC including $y=0.3$, $0.1<x<0.9$ and $0.1<z<0.9$.
The upper panel displays the perturbative progression from NLO (blue dotted) to NNLO (red dot-dashed) and N$^3$LO (black solid).
The lower panel, which normalizes results to the central NNLO prediction, reveals the dramatic stabilization of the perturbative expansion. 
We observe a moderate correction from N$^3$LO, starting at about $0.5$\%  for $z$ at $0.1$ and rising to $+4$\% at $0.9$.
We also plot the N$^3$LO predictions with an alternative choice of $\Delta=0.08$, which are consistent with those from our default choice of $\Delta=0.1$.
The scale uncertainty bands (hatched regions), calculated by varying $\mu$ by a factor of 2 from the nominal value $Q$, undergo a significant reduction when moving from NLO to N$^3$LO, confirming that the missing higher-order terms are now under control. 
In the lower panel, the diamond markers represent approximated N$^3$LO predictions obtained by adding the $\mathcal{O}(\alpha_s^3)$ threshold logarithmic contributions to the exact NNLO predictions.
These threshold corrections are calculated in Refs.~\cite{Abele:2022wuy, Goyal:2025bzf} including both leading and partial next-to-leading singular terms.
The threshold $K$-factor exhibits a distinct kinematic dependence, rising noticeably at large $z$. 
We observe good agreement between our N$^3$LO corrections and the approximated ones at $z>0.5$, indicating the dominance of threshold corrections in the large-$z$ region.
Results at a lower-energy EIC-like benchmark, $\sqrt{s}=45~\mathrm{GeV}$~\cite{Bonino:2024wgg}, are provided in Supplemental Material~\cite{SupplementalMaterial} and show similar perturbative behavior.

\begin{figure}[t!]
    \includegraphics[width=0.9\linewidth]{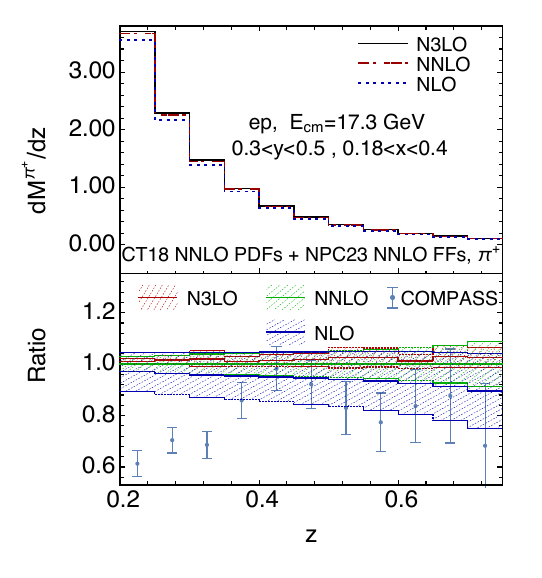}
    \vspace{-0.5cm}
    \caption{%
    {Differential multiplicity as a function of the hadron momentum fraction $z$ for charged pion production at the COMPASS experiment (proton target with $\sqrt s=17.3\,{\rm GeV}$) at various orders in QCD, including NLO, NNLO and N$^3$LO.
    In the lower panel, all predictions including scale variations are normalized to the central NNLO results.
    The error bars represent the COMPASS $\pi^+$ data with full uncertainties.}
    }
    \label{fig:compass}
\end{figure}
In Fig.~\ref{fig:x_dist} we show N$^3$LO predictions for the charged pion differential cross section as a function of the Bjorken variable $x$ for the same conditions as used in Fig.~\ref{fig:z_dist}.
We observe a very similar behavior of the N$^3$LO corrections compared to those in $z$ distributions for both the kinematic dependence and the size of corrections.
The N$^3$LO corrections remain positive throughout the considered $x$ range and reach about $+5\%$ at
$x=0.9$.
The predictions obtained with $\Delta=0.1$ and $0.08$ agree within the Monte Carlo uncertainties.
The N$^3$LO scale uncertainties are reduced to about half of those at NNLO in general, again indicating the missing higher-order terms are under control.
We also find a good agreement of our N$^3$LO corrections with the approximated ones for $x>0.4$.
Generalization of our method for calculating N$^3$LO corrections to Bjorken-$x$ distribution in polarized SIDIS is straightforward, which can be important for high-precision study of proton spin decomposition.
It would be interesting to see if similar perturbative convergence holds in polarized SIDIS. 
Finally, we present results for the COMPASS setup of muon scattering off a proton target, with $\sqrt{s}=17.3~\mathrm{GeV}$, $0.18<x<0.4$, and $0.3<y<0.5$.
Fig.~\ref{fig:compass} shows the differential multiplicity of $\pi^+$, which is the hadron production cross section divided by the inclusive deep-inelastic scattering (DIS) cross section, as a function of the hadron momentum fraction $z$, comparing to the COMPASS data~\cite{COMPASS:2024gje}.
We observe moderate size of N$^3$LO corrections and significant reduction of scale variations, which are now well below the experimental uncertainties.
On the other hand, the theoretical predictions overshoot the data consistently, especially in the low-$z$ region, indicating possible large uncertainties from the input of fragmentation functions~\cite{Gao:2024nkz,Gao:2024dbv}. 

\paragraph*{Summary and Outlook.---}
In conclusion we propose a novel method of two-dimensional transverse-momentum subtraction for calculations of identified hadron production applicable for both single-inclusive annihilation and deep-inelastic scattering.
We present the N$^3$LO QCD corrections to unpolarized SIDIS as a successful application, a first-ever result to this perturbative order for processes involving identified hadrons in the final state. 
The corrections are moderate in general, exhibiting excellent perturbative convergence and significantly reduced scale variations.
Our framework with fully differential descriptions allows for arbitrary selection cuts, and is especially applicable for efficient and precise study of hadron tomography at the upcoming Electron-Ion Collider.
Extension of the calculations to polarized SIDIS is within reach, which will further improve our understanding of the nucleon spin decomposition. 

\paragraph*{Acknowledgments.---}
The work of L.~D. and J.~G. is supported by the National Natural Science Foundation of China (NSFC) under Grant No.~12275173 and the Shanghai Municipal Education Commission under Grant No.~2024AIZD007.
S.~F. and D.~Y.~S. are supported by the National Natural Science Foundation of China under Grants No.~12675099, No.~12275052, No.~12147101, and No.~12547102, and the Innovation Program for Quantum Science and Technology under Grant No.~2024ZD0300101.
H.~T.~L. is supported by the National Natural Science Foundation of China under Grants No.~12275156 and No.~12321005, and the Shandong Provincial Department of Science and Technology under Project No.~TQ012025001.
H.~X.~Z. is supported by the National Natural Science Foundation of China under Grant No.~12425505.
The computations in this paper were run on the Siyuan-1 cluster supported by the Center for High Performance Computing at Shanghai Jiao Tong University.

\bibliographystyle{apsrev4-2}
\bibliography{SIDIS.bib}

\clearpage
\appendix
\onecolumngrid
\section*{Supplemental Material}

The solid cyan curves in Fig.~\ref{fig:cutoff_dep} of the main text represent fits to the genuine $\mathcal{O}({\alpha_s^3})$ corrections that include power corrections~\cite{Grazzini:2017mhc, Ebert:2019zkb}, which take the form of
\begin{align}
    \sigma^{(3)}(\Delta)=\Delta^2\left(a_5\ln^n(\Delta/4)+a_4\ln^{n-1}(\Delta/4)\right)+b_0\,,
\end{align}
where $n=5$ for the $uu$ channel, $n=4$ for $ug$, $gu$ and $n=3$ for $gg$, $ud$, $u\bar{u}$. 
Here we use three-parameter expressions containing the two leading power corrections. 
Given that the data points for different $\Delta$ are generated via independent Monte Carlo integrations, the resulting statistical uncertainties are uncorrelated. 
Therefore, the fitting is conducted using a $\chi^2$ minimization based on these independent errors, which gives $\chi^2/N_{\rm d.o.f.} \lesssim 1$ in all six channels. 
Physically, the fitted constant $b_0$ represents the value of the ${\rm N^3LO}$ corrections in the limit where $\Delta \to 0$.
\begin{figure}[h!]
    \includegraphics[width=0.55\linewidth]{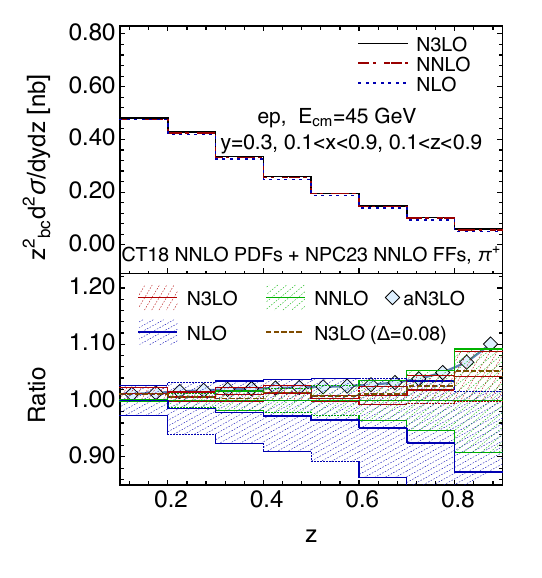}
    \caption{%
    Differential cross section (multiplied by the square of the bin center) as a function of the hadron momentum fraction $z$ for charged pion production at the EIC (45 GeV) at various orders in QCD, including NLO, NNLO, and N$^3$LO.
    In the lower panel, all predictions including scale variations are normalized to the central NNLO results.
    The brown dashed curve shows the N$^3$LO prediction obtained with $\Delta=0.08$; the default choice is $\Delta=0.1$.
    The diamond markers show the approximate N$^3$LO predictions.
    }
    \label{fig:az_dist}
\end{figure}
\begin{figure}[h!]
    \includegraphics[width=0.55\linewidth]{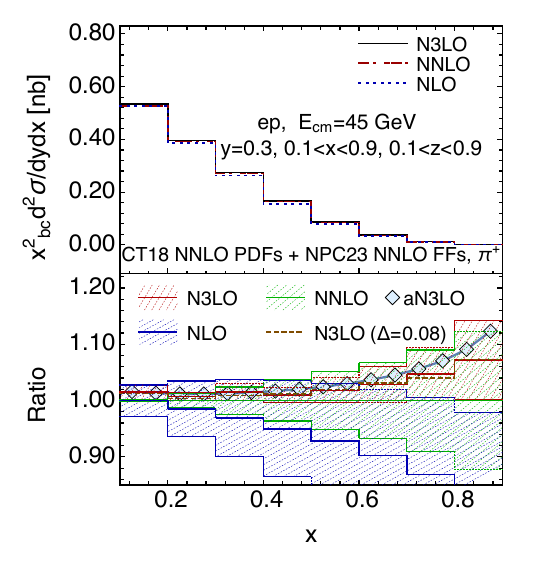}
    \caption{%
    The charged pion differential cross section at $\sqrt{s}=45~\mathrm{GeV}$ as a function of the Bjorken variable $x$.
    The perturbative orders, normalization, scale variations, slicing-cutoff choices, and approximate N$^3$LO markers follow the conventions of Fig.~\ref{fig:az_dist}.
}
    \label{fig:ax_dist}
\end{figure}
Figs.~\ref{fig:az_dist} and~\ref{fig:ax_dist} present the $z$- and $x$-differential cross sections, respectively, at $\sqrt{s}=45~\mathrm{GeV}$, with $y=0.3$, $0.1<x<0.9$, and $0.1<z<0.9$.
The upper panels display the perturbative progression from NLO (blue dotted) to NNLO (red dot-dashed) and N$^3$LO (black solid), while the lower panels are normalized to the central NNLO predictions.

The N$^3$LO correction to the $z$ distribution increases from about $+1\%$ at $z=0.1$ to $+5\%$ at $z=0.9$.
For the $x$ distribution, it increases from about $+1\%$ at $x=0.1$ to $+7\%$ at $x=0.9$.
The N$^3$LO predictions obtained with $\Delta=0.1$ and $0.08$ agree throughout the displayed ranges within the residual Monte Carlo uncertainties of approximately $0.5\%$.
The scale-variation bands are substantially reduced from NLO to N$^3$LO, showing perturbative behavior similar to that found at $\sqrt{s}=141~\mathrm{GeV}$.
In the lower panels, the diamond markers represent approximate N$^3$LO predictions obtained by adding the $\mathcal{O}(\alpha_s^3)$ threshold logarithmic contributions to the exact NNLO predictions.
These threshold contributions include the leading and partial next-to-leading singular terms calculated in Refs.~\cite{Abele:2022wuy,Goyal:2025bzf}.

\end{document}